\documentclass[final]{agujournal2019}
\usepackage{url} 
\usepackage{lineno}
\usepackage[inline]{trackchanges} 
\usepackage{soul}
\draftfalse

\journalname{Geophysical Research Letters}

\begin{document}

%
%


\title{Acute sensitivity of global ocean circulation and heat content to eddy energy dissipation time-scale}

%
%




\authors{J. Mak\affil{1,2,3}, D. P. Marshall\affil{1}, G. Madec\affil{4} and J. R. Maddision\affil{5}}


\affiliation{1}{Department of Physics, University of Oxford}
\affiliation{2}{Department of Ocean Science, Hong Kong University of Science and Technology}
\affiliation{3}{Center for Ocean Research in Hong Kong and Macau, Hong Kong University of Science and Technology}
\affiliation{4}{Sorbonne Universit\'es (University Pierre et Marie Curie Paris
6)-CNRS-IRD-MNHN, LOCEAN Laboratory, Paris}
\affiliation{5}{School of Mathematics and Maxwell Institute for Mathematical
Sciences, The University of Edinburgh}




\correspondingauthor{Julian Mak}{julian.c.l.mak@googlemail.com}




\begin{keypoints}
\item Key metrics of global ocean circulation (ACC transport, AMOC strength, OHC anomaly) acutely sensitive to eddy energy dissipation time-scale.
\item Modest variations in the dissipation time-scale has an comparable effect to significant variations in the Southern Ocean wind forcing.
\item Constraints on the dissipation time-scale critical to long-time integrations of ocean climate models such as paleoclimate scenarios.
\end{keypoints}

%
%

%
%


\begin{abstract}
The global ocean overturning circulation, critically dependent on the global
density stratification, plays a central role in regulating climate evolution.
While it is well-known that the global stratification profile exhibits a strong
dependence to Southern Ocean dynamics and in particular to wind and buoyancy
forcing, we demonstrate here that the stratification is also acutely sensitive
to the mesoscale eddy energy dissipation time-scale. Within the context of a
global ocean circulation model with an energy constrained mesoscale eddy
parameterization, it is shown that modest variations in the eddy energy
dissipation time-scale lead to significant variations in key metrics relating to
ocean circulation, namely the Antarctic Circumpolar Current transport, Atlantic
Meridional Overturning Circulation strength, and global ocean heat content, over
long time-scales. The results highlight a need to constrain uncertainties
associated with eddy energy dissipation for climate model projections over
centennial time-scales, but also for paleoclimate simulations over millennial
time-scales.
\end{abstract}


\section*{Plain Language Summary}
Modest uncertainties in mesoscale eddy energy dissipation time-scale translate
to significant variations in the global ocean circulation and heat content on
long time-scales. A 50\% change in the eddy energy dissipation time-scale has a
similar effect to halving and doubling of Southern Ocean wind stress in
quasi-equilibrium calculations, leading to a global ocean heat content change
that is an order of magnitude larger than those typically found in modern day
era reconstructions and projections. The results highlight a need to combine
theoretical, modeling, and observational efforts to constrain the uncertainties
in eddy energy dissipation for climate projections and paleoclimate
reconstructions.

%
%

%


%
%
%
%




\section{Overview and key findings}\label{sec:intro}

Evolution of the ocean stratification plays a fundamental role in climate
evolution, through the associated consequences for the global meridional
overturning circulation. Reconstructions of past climate together with the use
of numerical models have highlighted how shoaling and weakening of the Atlantic
Meridional Overturning Circulation (AMOC), associated with changes in the
deep/abyssal stratification, have important consequences for the global energy,
oxygen and carbon cycles \cite<e.g.,>{ZhangVallis13, Adkins13, Ferrari-et-al14,
Burke-et-al15, Bopp-et-al17, Jansen17, Takano-et-al18, GalbraithdeLavergne19}.
In particular, the Southern Ocean is ``\emph{disproportionately important}''
\cite{Newman-et-al19} for the global stratification profiles because of the
connection in the stratification profiles, with the implication that Southern
Ocean processes can exert a control on the global overturning circulation.

It is known that the Southern Ocean stratification is primarily dependent on
wind forcing \cite{ToggweilerSamuels95, Toggweiler-et-al06}, buoyancy forcing
\cite{Hogg10, Morrison-et-al11, Jansen17}, and to eddy dynamics
\cite{Munday-et-al13, Farneti-et-al15, Bishop-et-al16}. Focusing on mesoscale
eddies, an extra complication arises since there are notable divergences in
model response depending on how mesoscale eddies are represented, between
whether they are represented explicitly or parameterized \cite{Munday-et-al13,
Farneti-et-al15, Bishop-et-al16}, and the form of the parameterization
\cite{HofmanMoralesMaqueda11, ViebahnEden12, Meredith-et-al12, Munday-et-al13,
Farneti-et-al15, Bishop-et-al16}. While the issue of mesoscale eddy
representation ``\emph{frequently have a larger effect on ocean climate
sensitivity than the total effect of other classes of parameterizations}''
\cite{FoxKemper-et-al19}, there have been advances on the eddy parameterization
aspect, where the role of eddy energy in mesoscale eddy parameterizations is
increasingly being studied \cite{EdenGreatbatch08, MarshallAdcroft10,
Marshall-et-al12, Eden-et-al14}. Models with parameterized eddies employing eddy
energy constrained eddy diffusivities or transport coefficients display improved
model responses that are closer to the responses displayed in analogous high
resolution models \cite{JansenHeld14, Jansen-et-al15a, Jansen-et-al15b,
Mak-et-al17a, Mak-et-al18, Klower-et-al18, Bachman19}. In particular, the
GEOMETRIC parameterization \cite{Marshall-et-al12, Mak-et-al17a,
Marshall-et-al17, Mak-et-al18} --- effectively rescaling the standard
Gent--McWilliams \cite{GentMcWilliams90, Gent-et-al95} eddy transport
coefficient by the total eddy energy according to rigorous mathematical
identities \cite{Marshall-et-al12, MaddisonMarshall13} and supported in
diagnoses of eddy resolving calculations \cite{Bachman-et-al17} --- imparts an
Antarctic Circumpolar Current and Atlantic Meridional Overturning Circulation
sensitivity to changes in the Southern Ocean wind forcing to idealized ocean
climate models that are closer the analogous high resolution models
\cite{Mak-et-al18}.


A natural question to ask is how strong is the influence of mesoscale eddy
dynamics on the Southern Ocean as well as global ocean circulation. Since there
is a link between eddy energy and the degree of feedback arising from the
eddies, the mesoscale eddy energy dissipation time-scale should therefore play
an important role: if more energy is drained from the mesoscale eddy field, the
associated eddy form stress weakens, thereby reducing vertical momentum
transport, in turn modifying the momentum balance. Following this line of
argument, it was argued in \citeA{Marshall-et-al17} that the overall circumpolar
transport in the Southern Ocean should increase with increasing eddy energy
dissipation, and so, by thermal wind shear relation, lead to steeply tilting
isopycnals in the Southern and ocean and a deepening of the global pycnocline
depth over long time-scales.

The extent of the influence of eddy energy dissipation on the global circulation
is the primary focus of the present work. A key finding here is that a 50\%
variation around a control time-scale has a similar effect to halving and
doubling of the present day Southern Ocean wind stress on the modeled Antarctic
Circumpolar Current transport, Atlantic Meridional Overturning Circulation
strength, and the global integrated ocean heat content anomaly
(Fig.~\ref{fig:diag_diss} and Fig.~\ref{fig:diag_wind}), attributed primarily to
changes in the global pycnocline depth. While the Southern Ocean wind forcing is
not expected to vary so dramatically, the extent of plausible mesosacle eddy
energy dissipation time-scale is not known, due to a lack of theoretical and
observation constraints currently available. The results here thus highlight a
crucial need to combine theoretical, modeling, and observational efforts to
constrain the uncertainties in eddy energy dissipation, not only from a
theoretical point of view for understanding, but also for practical purposes in
constraining uncertain model parameters for numerical models used in climate
projections and paleoclimate reconstructions.


\section{Method and model description}\label{sec:method}

The principal focus here is on quasi-equilibrium sensitivities of the global
overturning circulation to the eddy energy dissipation time-scale. While one
might consider employing an eddy resolving ocean model for such a study, the
associated computational costs are prohibitive. Thus we employ a model with
parameterized eddies, and utilize the Nucleus for European Modelling of the
Ocean (NEMO, v3.7dev r8666) \cite{Madec-NEMO} in the global configuration (ORCA)
with realistic bathymetry, employing the tri-polar ORCA grid
\cite{MadecImbard96} and the LIM3 ice model \cite{Rousset-et-al15}. The present
ORCA1 model has a nominal horizontal resolution of $1^\circ$, employs 46 uneven
vertical levels, and is initialized with WOA13 climatology
\cite{Locarini-et-al13, Zweng-et-al13}. The model employs the TEOS-10 equation
of state \cite{Roquet-et-al15a}, with the atmospheric forcing modeled by the
NCAR bulk formulae with normal year forcing \cite{LargeYeager09}. Sea surface
salinity but not temperature restoration is included to reduce model drift.

An energetically constrained mesoscale eddy parameterization scheme is required,
and for our investigation the GEOMETRIC parameterization for mesoscale eddies
\cite{Marshall-et-al12, Mak-et-al18} was chosen and implemented in NEMO (see
Supplementary Information for implementation details). Briefly, GEOMETRIC
computes a horizontally and temporally varying coefficient for eddy induced
advection \cite{GentMcWilliams90, Gent-et-al95} according to \cite<cf. Eq. 4
of>{Mak-et-al18}
\begin{equation}\label{eq:kgm}
  \kappa_{\rm gm} = \alpha \frac{\int E\; \mathrm{d}z}{\int (M^2 / N)\; \mathrm{d}z},
\end{equation}
where $M$ and $N$ are the horizontal and vertical buoyancy frequencies, $\alpha$
is a non-dimensional tuning parameter (bounded in magnitude by 1), and $E$ is
the total (potential and kinetic) eddy energy. The depth-integrated eddy energy
$\int E\; \mathrm{d}z$ is provided by a parameterized eddy energy budget given
by \cite<cf. Eq. 2 of>{Mak-et-al18}
\begin{equation}\label{eq:ene-eq}
  \frac{\mathrm{d}}{\mathrm{d}t}\int E\; \mathrm{d}z
    + \underbrace{\nabla_H \cdot \left( \left(\widetilde{\mathbf{u}}^z - |c|\, \mathbf{e}_x\right) \int E\; \mathrm{d}z \right)}_\textnormal{advection}
   = \underbrace{\int \kappa_{\rm gm} \frac{M^4}{N^2}\; \mathrm{d}z}_\textnormal{source}
    - \underbrace{\lambda \int (E - E_0)\; \mathrm{d}z}_\textnormal{dissipation} 
    + \underbrace{\eta_E\nabla^2_H  \int E\; \mathrm{d}z}_\textnormal{diffusion}.
\end{equation}
The depth-integrated eddy energy is advected by the depth average flow
$\widetilde{\mathbf{u}}^z$ and propagated westward at the long Rossby wave phase
speed $|c|$ \cite{Chelton-et-al11, KlockerMarshall14}, has growth arising from
slumping of mean density surfaces, and diffused in the horizontal
\cite{Grooms15, Ni-et-al20a, Ni-et-al20b}, with $\nabla_H$ denoting the
horizontal gradient operator and the $\eta_E$ the associated eddy energy
diffusivity. A linear dissipation of eddy energy at rate $\lambda$ (but
maintaining a minimum eddy energy level $E_0$) is utilized, so $\lambda^{-1}$ is
the eddy energy dissipation time-scale. For this work, $\alpha = 0.04$ is
prescribed, partially informed by the results of \citeA{Poulsen-et-al19}, and
$\eta_E = 500\ \mathrm{m}^2\ \mathrm{s}^{-1}$ was chosen. While the
Gent--McWilliams coefficient follows the prescription given in Eq.~\ref{eq:kgm},
the isoneutral diffusion coefficient \cite{Griffies98} is kept constant at
$1000\ \mathrm{m}^2\ \mathrm{s}^{-1}$. Values of the isopycnal slopes used to
compute the parameterized eddy energy, eddy induced advection and isoneutral
diffusion are limited to $1/100$ in the interior, and linearly decreased from
the base of the model mixed layer to zero at the surface to maintain no flux
conditions. For more details of the GEOMETRIC parameterization and its
implementation in NEMO, please see the Supporting Information.

Given the lack of constraints on the values and uncertainties associated with
the eddy energy dissipation time-scale, for simplicity we take the eddy energy
dissipation time-scale $\lambda^{-1}$ to be a constant in space and time, with a
control value of $\lambda^{-1} = 100$ days \cite<e.g.,>[Marshall \& Zhai, pers.
comm.]{Melet-et-al15}, and consider a range of values spanning approximately
50\% about the control value (in this case six experiments with $\lambda^{-1}$
ranging 60 to 160 days in increments of 20 days). A control model utilizing
$\lambda^{-1} = 100$ days was first spun up for 1500 years, after which the
perturbation experiments were integrated for a further 1600 years; see
Supplementary Information Fig.~S1-4 for some of the resulting climatology. All
metrics presented in this work were diagnosed from data averaged over the last
100 model years. 


\section{Results}


\subsection{Sensitivity to eddy energy dissipation time-scale}

\begin{figure*}
\begin{center}
  \includegraphics[width=0.95\linewidth]{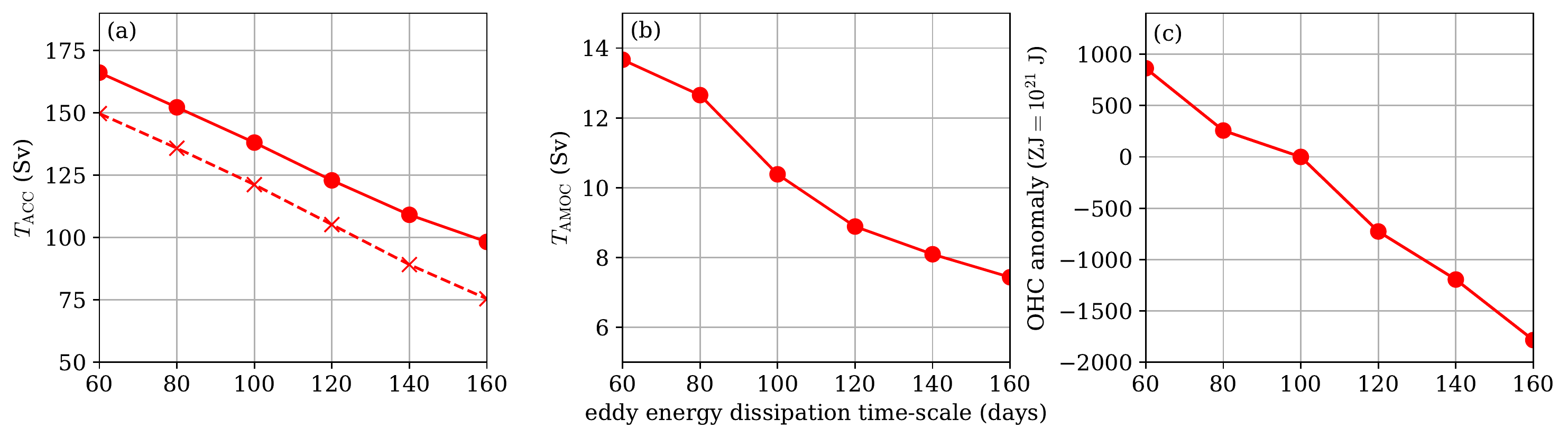}
  \caption{Diagnostics from the varying eddy energy dissipation time-scale
  experiments. Diagnostics are: ($a$) ACC transport (total in solid lines,
  thermal wind component in dashed lines); ($b$) AMOC strength; ($c$)
  domain-integrated ocean heat content anomalies as solid lines, where the
  anomalies are relative to the control calculation (one times wind
  amplification and dissipation time-scale of 100 days) with the value of 21,300
  ZJ.}
  \label{fig:diag_diss}
\end{center}
\end{figure*}

The key metrics of interest here are the total Antarctic Circumpolar Current
(ACC) transport, Atlantic Meridional Overturning Circulation (AMOC) strength,
and the globally integrated Ocean Heat Content (OHC) anomaly relative to the
control calculation, respectively given as the transport through the model Drake
passage, the transport over the top $1000\ \mathrm{m}$ at the model 26$^\circ$ N
on the Western side of the Atlantic, and the global integrated conservative
temperature multiplied accordingly by the density and heat capacity.
Fig.~\ref{fig:diag_diss} compares these metrics diagnosed from experiments
varying the eddy energy dissipation time-scale. Increasing the dissipation
time-scale (i.e., decreased damping of the eddies) leads to a substantial
decrease in the ACC transport, AMOC strength, and total ocean heat content
anomaly, which can be attributed to the deepening of the global pycnocline,
consistent with theoretical arguments \cite{Marshall-et-al17}. In particular, we
note that the changes in the OHC anomalies found in post-industrial period
reconstructions \cite{Levitus-et-al12, Cheng-et-al17, Cheng-et-al19,
Zanna-et-al19} are typically on the order of $10^{23}\ \mathrm{J}$ ($100\
\mathrm{ZJ}$), while the changes to total OHC associated with the uncertainties
in eddy energy dissipation time-scale here can be an order of magnitude larger
($10^{24}\ \mathrm{J}$).

\begin{figure}
\begin{center}
  \includegraphics[width=\linewidth]{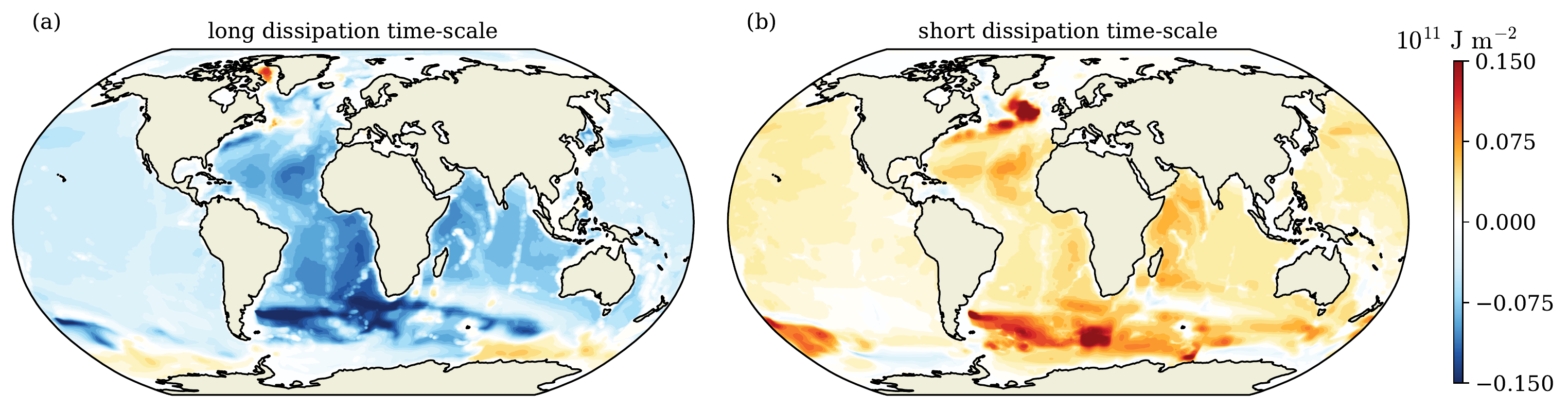}
  \caption{Depth-integrated ocean heat content anomaly (relative to the control
  calculation with dissipation time-scale of 100 days and one times wind
  amplification) for varying dissipation experiments ($a,b$, at $\lambda^{-1} =
  $ 160 and 60 days respectively).}
  \label{fig:heatc_comparison}
\end{center}
\end{figure}

The distribution of the lateral depth-integrated ocean heat content for varying
eddy energy dissipation time-scale is shown in Fig.~\ref{fig:heatc_comparison}.
Varying the eddy energy dissipation time-scale leads to a significant global
change in the OHC anomalies, attributed mainly to the changes in pycnocline depth
(Supporting Information, Fig.~S6-9). Note also that the changes appear to be
most significant over the Southern Ocean and in the Atlantic basin, attributed
to significant changes in the AMOC as well as the overturning within the
Southern Ocean (Supporting Information, Fig.~S3-4).

Here, changing the eddy energy dissipation time-scale $\lambda^{-1}$ affects the
total eddy energy $E$, which in turn impacts the Gent--McWilliams coefficient
$\kappa_{\rm gm}$. While the significant changes to global OHC and circulation
arising from changing $\kappa_{\rm gm}$ has been noted before
\cite<e.g.>{ZhangVallis13}, the fundamental difference here is that the
sensitivities are arising through uncertainties in the eddy energy dissipation
that happens to impact the Gent--McWilliams parameter, and the eddy energy
dissipation is a process that in principle is perhaps more amenable to be
constrained by theoretical, numerical or observational means.


\subsection{Sensitivity to Southern Ocean wind forcing}

For completeness, experiments varying Southern Ocean wind forcing were also
performed. The zonal wind stress over the Southern Ocean region within the model
is amplified instead of the imposed zonal wind speed, so that any modifications
to the ocean surface evaporation and turbulent fluxes as calculated through the
bulk formulae occurs through changes to the ocean state rather than the imposed
wind forcing. Two sets of perturbation experiments were performed: ($a$) a
$\kappa_{\rm gm}$ that is varying in the horizontal and in time as given by
Eq.~\ref{eq:kgm}, with no further re-tuning (denoted GEOM), and ($b$) a
prescribed $\kappa_{\rm gm}$ diagnosed from the last 100 years of the control
spin-up (denoted GM), but one that is now time-independent although still
spatially varying.

\begin{figure*}
\begin{center}
  \includegraphics[width=0.95\linewidth]{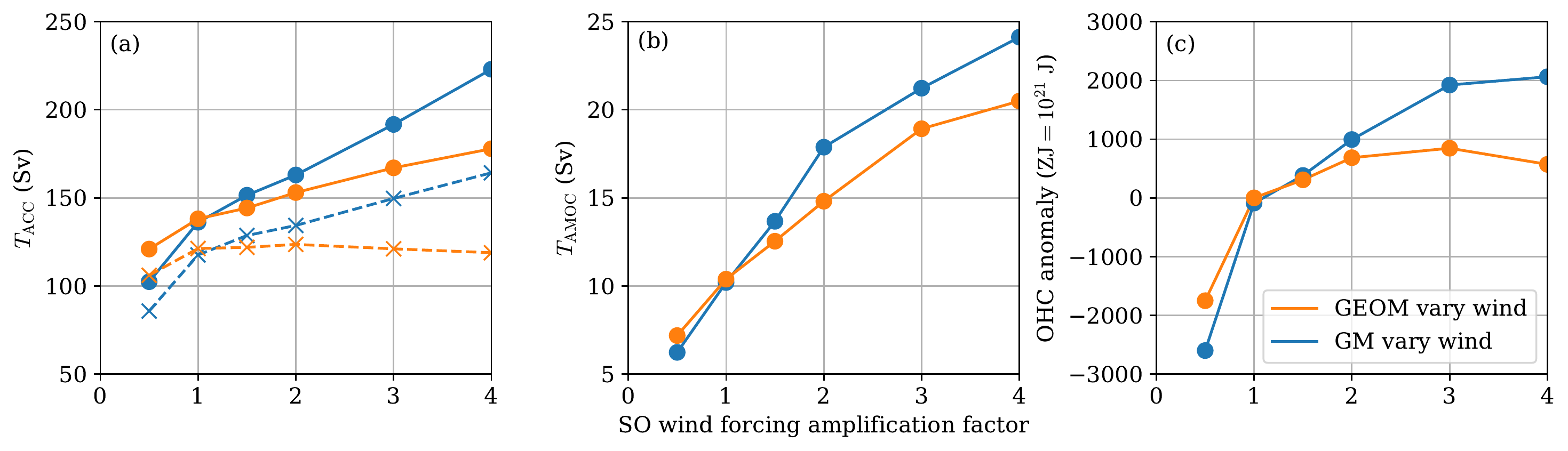}
  \caption{Diagnostics from the varying Southern Ocean wind stress experiments
  (with GEOM and GM calculations in orange and blue respectively), showing:
  ($a$) ACC transport (total in solid lines, thermal wind component in dashed
  lines); ($b$) AMOC strength; ($c$) domain-integrated ocean heat content
  anomaly, where the anomalies are relative to the respective control
  calculations (one times wind amplification and dissipation time-scale of 100
  days) with the value of 21,300 ZJ.}
  \label{fig:diag_wind}
\end{center}
\end{figure*}

Fig.~\ref{fig:diag_wind} shows the sensitivities of the same global ocean
climatological metrics to changes in the imposed Southern Ocean wind forcing. In
Fig.~\ref{fig:diag_wind}($a$), while the total ACC transport (solid lines)
increases with wind forcing, the ACC thermal wind transport (dashed lines) in
the GEOM calculations (orange lines, cross markers) is relatively insensitive to
changes in the wind forcing, demonstrating the eddy saturation phenomenon
\cite{HallbergGnanadesikan06, MeredithHogg06, Munday-et-al13, Farneti-et-al15,
Bishop-et-al16}. On the other hand, the corresponding AMOC strength shown in
Fig.~\ref{fig:diag_wind}($b$) in the GEOM calculations display a reduced
sensitivity to changes in the Southern Ocean wind forcing relative to the GM
case, and is related to the phenomenon of eddy compensation
\cite{GentDanabasoglu11, HofmanMoralesMaqueda11, ViebahnEden12,
Meredith-et-al12, Munday-et-al13, Farneti-et-al15, Bishop-et-al16}. The
aforementioned sensitivities coincide with a weaker sensitivity of the global
pycnocline depth to changes in Southern Ocean wind forcing
\cite{Marshall-et-al17, Mak-et-al18}. Comparing Fig.~\ref{fig:diag_wind} with
Fig.~\ref{fig:diag_diss}, a 50\% change about a chosen eddy energy dissipation
time-scale has a comparable effect to halving or doubling the Southern Ocean
wind forcing in the diagnosed metrics. As with the varying eddy energy
dissipation experiments, the OHC anomalies are particularly significant over the
Southern Ocean and the Atlantic basin (Supporting Information, Fig.~S5). In the
present case of varying wind stress, however, the notable variations in the OHC
anomalies are attributed to significant changes in the abyssal watermass
properties (Supporting Information, Fig.~S6-9). The observed changes in the
watermass properties may perhaps be attributed to a modified sea ice extent
(Supporting Information, Fig.~S10) via changes in the sea ice export by the
wind, leading to changes in deep water formation and abyssal watermass
properties, analogous to the mechanism proposed in \citeA{Ferrari-et-al14} and
\citeA{Burke-et-al15}.

\section{Summary and outlooks}

The present work demonstrates that, within the context of a global configuration
ocean model with an energetically constrained mesoscale eddy parameterization,
modest and perhaps not implausible variations in the mesoscale eddy energy
dissipation time-scale translate to significant sensitivities of the diagnosed
Antarctic Circumpolar Current transport, Atlantic Meridional Overturning
Circulation strength, and the global Ocean Heat Content over long time-scales in
the modeled ocean. The physical reasons for the sensitivity is that modifying
the eddy energy dissipation leads to changes in the mesoscale eddy dynamics in
the Southern Ocean, that in turn lead to significant changes to the global ocean
stratification over long time-scales. The sensitivity of the aforementioned key
ocean climatological metrics to eddy energy dissipation time-scale is found to
be comparable to those found for varying Southern Ocean wind forcing, where a
50\% change about a chosen eddy energy dissipation time-scale has a comparable
effect to halving or doubling the Southern Ocean wind forcing in the diagnosed
ocean circulation metrics. In particular, changes to the globally integrated
Ocean Heat Content anomalies can vary by up to an order of magnitude larger than
for reconstructions for total ocean heat content for the anthropogenic period
\cite{Levitus-et-al12, Cheng-et-al17, Cheng-et-al19, Zanna-et-al19}, and
comparable to the end of 21$^\textnormal{st}$ projections under the
Representative Concentration Pathways scenarios \cite<see figure
in>{Cheng-et-al19}. While the changes in the Southern Ocean wind forcing are not
expected to vary to the extent considered in this work \cite<e.g.>{Lin-et-al18},
there are no strong theoretical, numerical or observational constraints on the
eddy energy dissipation time-scale and its distribution (but see next paragraph
on works towards constraining the energy fluxes of the contributing processes).
There is thus a need to combine and dedicate theoretical, modeling and
observational efforts to constrain the uncertainties in the eddy energy
dissipation time-scale, given the impact the associated uncertainties can have.

\begin{figure}
\begin{center}
  \includegraphics[width=\linewidth]{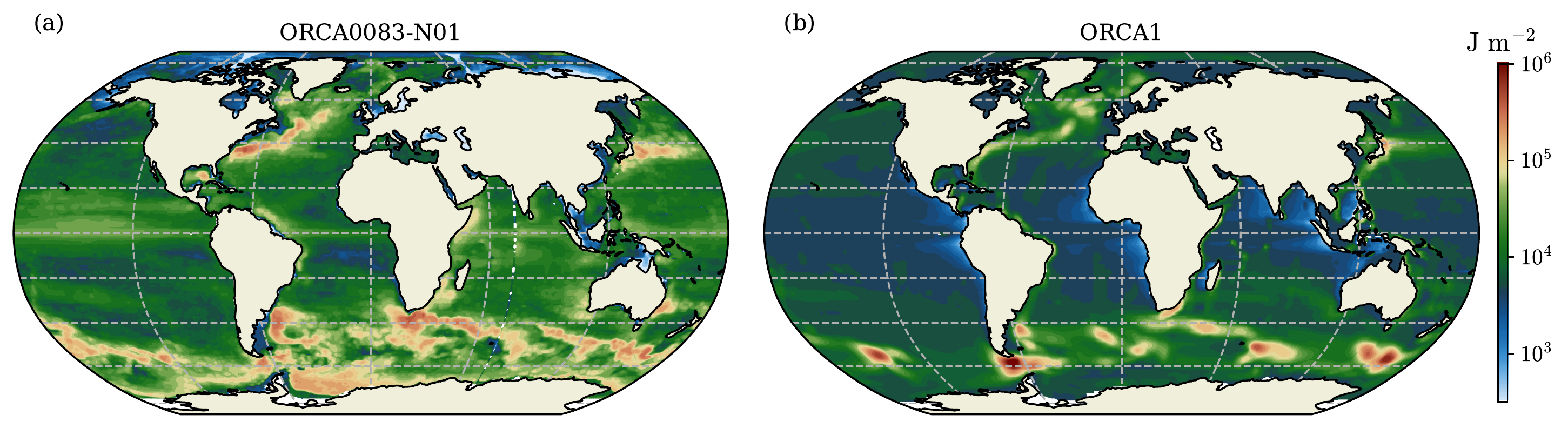}
  \caption{Depth-integrated total (kinetic and potential) eddy energy density
  (in units of $\mathrm{J}\ \mathrm{m}^{-2}$), diagnosed from ($a$) the high
  resolution ORCA0083-N01 calculation with explicit eddies, and ($b$) from the
  GEOM control calculation.}
  \label{fig:energy}
\end{center}
\end{figure}

In the present work the eddy energy dissipation is linear \cite<cf.,>{Klymak18}
with a time-scale that is a prescribed constant in space and time. One
particular consequence of a prescribed spatially constant eddy energy
dissipation time-scale may be seen in Fig.~\ref{fig:energy}, which shows the
total (kinetic and potential) eddy energy diagnosed from a high-resolution
global configuration model and from the control experiment here. While the eddy
energy signature displays some similarities in terms of spatial patterns in the
Southern Ocean and Western Boundary Current regions, there is clearly room for
improvement for the parameterized case. For example, the eddy energy signature
in the parameterized case is too weak in the Western Boundary Currents and in
the equatorial region, attributed to the fact that the spatially constant eddy
energy dissipation time-scale was chosen somewhat with the Southern Ocean in
mind, and is probably too short for the ocean basin regions (Marshall \& Zhai,
pers. comm.). The mesoscale eddy energy dissipation time-scale is expected to be
a more complicated function than the choice taken here and, fundamentally,
should depend on a wide variety of dynamical processes such as bottom drag
\cite<e.g.>{Sen-et-al08, Ruan-et-al21}, non-propagating form drag
\cite{Klymak18, Klymak-et-al21}, return to mean-flow \cite<e.g.>{Bachman19,
Jansen-et-al19} scattering into internal waves \cite<e.g.>{Nikurashi-et-al13,
Melet-et-al15, MacKinnon-et-al17, Yang-et-al18, Sutherland-et-al19}, loss of
balance \cite<e.g.>{Molemaker-et-al05, Barkan-et-al17, Rocha-et-al18}, and eddy
killing by the wind \cite<e.g.>{Xu-et-al16, Rai-et-al21}. Though the challenges
in constraining the uncertainties in the eddy energy dissipation time-scale are
formidable, the observed/diagnosed eddy energy signature can perhaps act as a
target towards efforts to constrain the aforementioned unknowns, highlighting
the potential for further research relating to ocean energetic pathways and its
consequences for climate evolution (see for example \citeA{Ruan-et-al21} for a
recent review of research relating to ocean eddy energy pathways).

While the present results are based on the choice of utilizing the GEOMETRIC
parameterization for mesoscale eddies \cite{Mak-et-al18}, given the link of the
eddy energy dissipation time-scale and the resulting eddy induced circulation
through the eddy energy (in this work through the Gent--McWilliams coefficient
for eddy induced advection), the sensitivities of key ocean climatological
metrics to the eddy energy dissipation is expected to carry over if other eddy
energy based parameterization schemes for mesoscale eddies
\cite<e.g.>{Jansen-et-al15b, Bachman19} are utilized, or if analogous
experiments are carried out in global eddy resolving models varying possible
mechanisms of mesoscale eddy energy dissipation \cite<e.g. bottom drag,
cf.>{Marshall-et-al17}, although the magnitude of the sensitivities may differ.
While the present work focused on quasi-equilibrium calculations, similar
conclusions but with reduced magnitudes of the sensitivities are expected for
centennial time-scale calculations. What is clear, however, is that the present
work has significant consequences for paleoclimate simulations involving the
ocean, such as Paleoclimate Modelling Intercomparison Project calculations
\cite<PMIP,>{Kageyama-et-al18}, given the long time-scales inherently required
for the related simulations. Potential impact assessment for climate projections
and paleoclimate simulations in light of the present work will be investigated
and reported in due course.

\acknowledgments

This work was funded by the UK Natural Environment Research Council grant
NE/R000999/1 and utilized the ARCHER UK National Supercomputing Service. JM also
acknowledges financial support from the RGC Early Career Scheme 2630020 and the
Center for Ocean Research in Hong Kong and Macau, a joint research center
between the Qingdao National Laboratory for Marine Science and Technology and
Hong Kong University of Science and Technology. The data used for generating the
plots in this article is available through
\verb|http://dx.doi.org/10.5281/zenodo.5732755|. The authors would like to thank Andrew
Coward for providing access to the ORCA0083-N01 dataset through the ARCHER RDF
service, Xiaoming Zhai for discussions relating to eddy energy data, and George
Nurser for discussions in relation to the GEOMETRIC implementation and outlooks.


%
%

\bibliography{refs}

%
%
%
%
%

\end{document}